\documentclass[preprint,3p,times]{elsarticle}

\usepackage{lineno,hyperref}
\usepackage[T1]{fontenc}
\usepackage{comment}
\usepackage{epstopdf}
\usepackage{amsmath}
\usepackage{amssymb}
\usepackage{graphicx}
\usepackage{epsfig}
\usepackage{latexsym}
\usepackage{pstricks}
\usepackage{pst-plot}
\usepackage{amsmath, amsthm, amsfonts, amssymb}
\usepackage{appendix}
\usepackage[bitstream-charter]{mathdesign}
\modulolinenumbers[5]

\journal{Journal of \LaTeX\ Templates}

\begin{document}

\begin{frontmatter}

\title{Notions of the ergodic hierarchy for curved statistical manifolds}
\author[iflp]{Ignacio S. Gomez\corref{cor1}}
\ead{nachosky@fisica.unlp.edu.ar}
\cortext[cor1]{Corresponding author}
\address[iflp]{IFLP, UNLP, CONICET, Facultad de Ciencias Exactas, Calle 115 y 49, 1900 La Plata, Argentina}

\begin{abstract}
We present an extension of the ergodic, mixing, and Bernoulli levels of the ergodic hierarchy for statistical models on curved manifolds, making use of elements of the information geometry. This extension focuses on the notion of statistical independence between the microscopical variables of the system.
Moreover, we establish an intimately relationship between statistical models and family of probability distributions belonging to the canonical ensemble, which for the case of the quadratic Hamiltonian systems provides a closed form for the correlations between the microvariables in terms of the temperature of the heat bath as a power law. From this we obtain an information geometric method for studying Hamiltonian dynamics in the canonical ensemble. We illustrate the results with two examples: a pair of interacting harmonic oscillators presenting phase transitions and the $2\times 2$ Gaussian ensembles. In both examples the scalar curvature results a global indicator of the dynamics.
\end{abstract}


\end{frontmatter}

\nolinenumbers

\section{Introduction}
The possibility of that relevant features of the dynamics of a system could be obtained from the differential geometric structure of the probabilities distributions gave place to the first encounter between differential geometry and probability theory.
In this sense, geometrization of thermodynamics and statistical mechanics constituted the most important achievement in the subject with several approaches
like the obtained by means of the internal energy as considered Weinhold \cite{weinhold}, or the Ruppeiner metric given by the second moments of thermodynamical fluctuations \cite{ruppeiner}, among others. With the same Riemannian character of these approaches, another formulations based in the thermodynamic of parameters were established in the field of the statistical mechanics like the foundational works of Rao \cite{rao} and Amari \cite{amari}, and also given by Ingarden \cite{ingarden}, Janiszek \cite{janyszek}.
The successful application of all this vast body of approaches in characterizing several phenomena, such as the phase transitions and their critical points in non ideal gases, gave them entity to constitute a discipline within the information theory, called \emph{Information Geometry}.
Curved statistical manifolds are the subject of study of the information geometry, and they have associated the Fisher--Rao metric \cite{rao} which in turn is linked to the concepts of entropy and Fisher information. Generalized extensions of the information geometry \cite{abe,naudts,portesi} with regard to nonextensive formulation of statistical mechanics \cite{tsallis} has been also considered.
The utility of information geometry is not only limited to thermodynamics and statistical mechanics. For instance, it has been applied in quantum mechanics leading a quantum generalization of the Fisher metric \cite{bures}, and recently also in nuclear plasmas \cite{geert}.
In particular, an application of information geometry to chaos can be performed by considering complexity on curved manifolds \cite{cafaro,cafaro2,cafaro3,cafaro4}. In this approach, asymptotic expressions for information measures are obtained by means of geodesic equations leading to a \emph{criterion} for characterizing global chaos on statistical manifolds \cite{cafaro4}: the more negative is the curvature, the more chaotic is the dynamics. As usual, chaos can be characterized in terms of diverging initially nearby trajectories \cite{lich}. For the statistical models this condition results in the divergence of geodesic paths on the statistical manifold and constitutes a local criterion for chaos.

Besides, in dynamical systems theory, the \emph{ergodic hierarchy} (EH) characterizes the chaotic behavior in terms of a type of correlation between subsets of the phase space \cite{berko,lasota}. In the asymptotic limit of large times, the EH establishes that the dynamics is more chaotic when the correlation decays faster.
According to correlation decay, the four levels of EH are, from the weakest to the strongest: ergodic, mixing, Kolmogorov, and Bernoulli.
In particular, in mixing systems any two subsets enough separated in time can be considered as ``statistically independent" which allows one to use a statistical description of the behavior of the system.
In quantum chaos, the statistical independence is present in the universal statistical properties of energy levels which are given by the Gaussian ensembles \cite{stockmann,haake,casati libro,casati}. In Gaussian ensembles theory one assumes that in a fully chaotic quantum system the interactions are neglected in such way that the Hamiltonian matrix elements can be considered statistically independent \cite{bgs}. Related to this, in \cite{paper1,paper2,paper3} a quantum extension of the EH was proposed, called the \emph{quantum ergodic hierarchy}, which allowed to provide a characterization of the chaotic behaviors of the Casati--Prosen model \cite{casati} and the kicked rotator \cite{stockmann,haake, casati libro}.

Inspired by the characterizations of quantum chaotic systems made in \cite{paper1,paper2,paper3,paper4,paper5} and making use of curved statistical models, we propose a generalization of the ergodic, mixing and Bernoulli levels of the EH in the context of the information geometry, which we called \emph{Information Geometric Ergodic Hierarchy} (IGEH). In order to use it, we define a distinguishability measure for a $2D$ correlated model that allows us to give an upper bound for the correlation of IGEH. Moreover, considering Hamiltonian systems belonging to the canonical ensemble we also give a method for characterizing their dynamics in terms of the statistical parameters and the levels of the IGEH.


In this way, our main contribution is two--fold: 1) an intimately connection between statistical models and probability distributions of the canonical ensemble which allows one to geometrize the phase transitions, and 2) an information geometric version of the ergodic hierarchy as an alternative framework for studying the chaotic dynamics in curved statistical models.

The paper is organized as follows. In Section \ref{sec prelim}, we give the notions and concepts of information geometry used throughout the paper, along with brief description of a $2D$ correlated model. In Section \ref{sec EH}, we make a brief review of the levels the ergodic hierarchy. Section \ref{sec IGEH} is devoted to an information geometric definition of the ergodic hierarchy by expressing the correlations in terms of probability distributions instead of subsets of phase space. Next, in Section \ref{sec 2Dmeasure} we define a distinguishability measure for the $2D$ correlated model and an upper bound for the correlation of the IGEH is given.
In Section \ref{sec geometrical canonical}, we establish the connection between the statistical models and the family of probability distributions belonging to the canonical ensemble. For quadratic Hamiltonian systems we show that their associated statistical models are the multivariate Gaussian ones, for which we obtain a closed form of the determinant of the covariance matrix in terms of the Hessian of the Hamiltonian. Here we also give an upper bound for the IG correlation where the temperature of the heat bath is considered as an external parameter.
In Section \ref{sec relevance}, we illustrate the formalism with two examples: a pair of interacting harmonic oscillators presenting phase transitions in the canonical ensemble and the $2\times2$ Gaussian Orthogonal Ensemble (GOE). Also, a panoramic outlook of the IGEH is sketched. Finally, in Section \ref{sec conclusions} we draw some conclusions, and future research directions are outlined.

\section{Elements of information geometry}\label{sec prelim}

We begin by introducing some fundamentals and concepts, following the definitions given in \cite{amari}.

\subsection{Statistical models}\label{subsec statistical models}
Given an abstract set $X$ one can consider the set $\textit{M}$ of all the probability density functions (PDFs) $p$
defined on $X$, i.e.
\begin{eqnarray}\label{2-1}
\textit{M}=\left\{ \ p \ : \ p:X\rightarrow\mathbb{R}\ , \ p(x)\geq0 \ , \ \int_X p(x)dx \ \right\}
\end{eqnarray}
where $\mathbb{R}$ is the set of real numbers and the integration must be replaced by a sum when $X$ is discrete. Consider a subset $S\subset \textit{M}$ such that each element of $S$ may be parameterized using a $m$--real vector $\theta=(\theta_1,\ldots,\theta_m)$ so that
\begin{eqnarray}\label{2-2}
S=\left\{ \ p_{\theta}\in \textit{M} \ : p_{\theta}=p(x;\theta) \ , \ \theta=(\theta_1,\ldots,\theta_m)\in \Theta \ \right\}
\end{eqnarray}
where $\Theta$ is a subset of $\mathbb{R}^m$. Then, if the mapping $\theta\mapsto p_{\theta}$ is injective it is said that $S$ is a \emph{statistical model} on $X$. The dimension of the statistical model is that of the macrospace, i.e. it is $m$--dimensional.
The physical interpretation of $X$ and $\Theta$ is as follows. Generally, $X$ represents the microscopic variables of the system under study which typically are difficult to control, for instance the positions of all the particles in a gas. Thus, $X$ is called the \emph{microspace} and $x$ are the microvariables. On the other hand,
$\Theta$ represent the macroscopic variables that can be measured in an experiment, like the mean value or the moments of the microvariables. It is said that $\Theta$ is the \emph{macrospace} and $\theta_1,\ldots,\theta_m$ are the macrovariables.
Since the microspace is fixed by the system then one can only choose the macrospace, and in this way the statistical model is established. Then, the statistical models are system--specific, from which follows that a statistical model could be useful for a system while that for another not.
\subsection{Metric structure of the statistical manifold}\label{subsec statistical manifold}
Next step is to describe the behavior of a system by means of a previously and adequately chosen, statistical model. For this, some kind of dynamics must be introduced on the statistical model. In information geometry this is accomplished by means of the Fisher--Rao tensor
\begin{eqnarray}\label{fisherao}
g_{ij}(\theta)=\int_X dx \ p(x;\theta)\frac{\partial \log p(x;\theta)}{\partial \theta_i }\frac{\partial \log p(x;\theta)}{\partial \theta_j }  \ \ \ \ \ \ \ \ \ \ \ \ \ \ \ i,j=1,\ldots,m
\end{eqnarray}
where $p(x;\theta)$ is a generic element of $S$. The metric tensor $g_{ij}$ endows the dynamics to the macrospace in terms of the geodesic equations for the macrovariables $\theta_i$, i.e. $S$ results to be a statistical manifold. More precisely, $S$ is a Riemmanian manifold and the statistical character is due to the elements of $S$ are probability distributions.

Thus, the main goal of the statistical models is to obtain some relevant information about the dynamics by means of the geodesic equations and geometrical quantities like the Ricci tensor, the scalar curvature etc. In this sense, we will use a criteria given by Cafaro et al. \cite{cafaro,cafaro2,cafaro3} to characterize global chaos on statistical models: \emph{``the more negative is the scalar curvature, the more chaotic is the dynamics"}.
From the metric tensor \eqref{fisherao} one can obtain the geodesic equations for the macrovariales $\theta_1,\ldots,\theta_m$ along with the following geometrical quantities that we will use throughout the paper.
\begin{eqnarray}\label{geodesic}
\textrm{Geodesic equations}: \ \ \ \ \ \ \frac{d\theta_k}{d\tau}+\Gamma_{ij}^k\frac{d\theta_i}{d\tau}\frac{d\theta_j}{d\tau}= \ \ \ , \ \ \ \forall \ k=1,\ldots,m
\end{eqnarray}
\begin{eqnarray}\label{christoffel}
\textrm{Christoffel symbols}: \ \ \ \ \ \ \Gamma_{ij}^k=\frac{1}{2}g^{im}\left(g_{mk,l}+g_{ml,k}-g_{kl,m} \right)
\end{eqnarray}
\begin{eqnarray}\label{riemann}
\textrm{Riemman curvature tensor}: \ \ \ \ \ \ R_{iklm}=\frac{1}{2}\left(g_{im,kl}+g_{kl,im}-g_{il,km}-g_{km,il}\right)+g_{np}\left(\Gamma_{kl}^n\Gamma_{im}^p-\Gamma_{km}^n\Gamma_{il}^p \right)
\end{eqnarray}
\begin{eqnarray}\label{ricci}
\textrm{Ricci tensor}: \ \ \ \ \ \ R_{ik}=g^{lm}R_{limk}
\end{eqnarray}
\begin{eqnarray}\label{scalar}
\textrm{Scalar curvature}: \ \ \ \ \ \ R=g^{ik}R_{ik}
\end{eqnarray}
where the comma in the subindexes denotes the partial derivative operation (of first and second orders), $g^{kl}$ is the inverse of $g_{ij}$, and $\tau$ is a parameter that characterizes the geodesic curves.

\subsection{The $2D$ correlated Gaussian model}\label{subsec 2Dmodel}

Most statistical models used in the literature are the so called \emph{Gaussian models}, due to its wide versatility for describing multiple phenomena. This models are obtained by choosing the subfamily $S$ as the set of multivariate Gaussian distributions. If $(x_1,\ldots,x_n)\in \mathbb{R}^n$ are the microvariables and there is no correlations between them then $(\mu_1,\ldots,\mu_n,\sigma_1,\ldots,\sigma_n)\in \mathbb{R}^n\times\mathbb{R}_+^n$ are the set of macrovariables, where $\mu_i$ and $\sigma_i^2$ correspond to the mean value and the variance of the $i$--th microvariable. However, for a more realistic description of the system the correlations between each of the microvariables must be taken into account. Considering the family of bivariate (binormal) distributions, one of the Gaussian models of lower dimension that present correlations can be obtained, which is given by
\begin{eqnarray}\label{bivariate}
p(x,y;\mu_x,\mu_y,\sigma_x,\sigma_y,r)=
\frac{1}{2\pi\sigma_x\sigma_y\sqrt{1-r^2}}\exp\left(-\frac{1}{2(1-r^2)}\left[\frac{(x-\mu_x)^2}{\sigma_x^2}+
\frac{(y-\mu_y)^2}{\sigma_y^2}-\frac{2r(x-\mu_x)(y-\mu_y)}{\sigma_x\sigma_y}\right]\right)
\end{eqnarray}
where $\sigma_{xy}=r\sigma_{x}\sigma_{y}$ is the covariance between $x$ and $y$ and $r$ is the correlation coefficient that assumes values within the ranges $-1\leq r\leq1$. Here the microspace is $X=\{(x,y)\in\mathbb{R}^2\}$ and the macrospace is $\Theta=\{(\mu_x,\mu_y,\sigma_x,\sigma_y)\in\mathbb{R}\times\mathbb{R}\times\mathbb{R}_{+}\times\mathbb{R}_{+}\}$. Nevertheless, one still can obtain a non trivial description by adding the following \emph{macroscopic constraint}
\begin{eqnarray}\label{constraint}
\Sigma^2=\sigma_x\sigma_y
\end{eqnarray}
where $\Sigma$ is a constant belonging to $\mathbb{R}_+$. Mathematically, the effect of $\Sigma$ is to restrict the dynamics to the submanifold $\Theta \cap \{\Sigma=\sigma_x\sigma_y\}$. Physically, $\Sigma$ resembles the minimum uncertainty relation when one chooses $x$ as the position of a particle and $y$ its conjugate variable. Moreover, this interpretation allows to give an explanation of the phenomenon of ``suppression of classical chaos
by quantization" from an information geometric point of view \cite{cafaro4}. For the sake of simplicity one also can fix the mean value $\mu_y$ of $y$ as zero. With the help of \eqref{constraint} then one can rewrite \eqref{bivariate}, thus obtaining the non--trivial correlated statistical model of lower dimensionality
\begin{eqnarray}\label{2Dmodel}
p(x,y;\mu_x,\sigma,r)=\frac{1}{2\pi\Sigma^2\sqrt{1-r^2}}\exp\left(-\frac{1}{2(1-r^2)}\left[\frac{(x-\mu_x)^2}{\sigma^2}+
\frac{y^2\sigma^2}{\Sigma^4}-\frac{2r(x-\mu_x)y}{\Sigma^2}\right]\right)
\end{eqnarray}
where we renamed $\sigma_x$ as $\sigma$. This is the so called \emph{the $2D$ correlated model} \cite{cafaro4}, since it present correlations between $x$ and $y$ by means of $\Sigma^2=\sigma_x\sigma_y$ and its macrospace $\Theta=\{(\mu_x,\sigma)\in \mathbb{R}\times\mathbb{R}_{+}\}$ is bidimensional.
It should be noted that here $r$ is considered as an external parameter that does not belong to the macrospace. For this model, from Eqs. \eqref{fisherao}--\eqref{scalar} one can obtain the Fisher tensor along with the following geometrical quantities.
\begin{eqnarray}\label{2Dmodelfisher}
\textrm{Fisher--Rao metric}: \ \ \ \ \ \ g_{ij}(\theta)=\left(
\begin{array}{cc}
\frac{1}{\sigma^2(1-r^2)} & 0 \\
0 & \frac{4}{\sigma^2(1-r^2)} \\
\end{array}
\right)
\end{eqnarray}
\begin{eqnarray}\label{2Dgeometrical}
\textrm{Non--vanishing Christoffel symbols}: \ \ \ \ \ \ \Gamma_{12}^1=\Gamma_{21}^1=-\frac{1}{\sigma} \ , \ \Gamma_{11}^2=\frac{1}{4\sigma} \ , \ \Gamma_{22}^2=-\frac{1}{\sigma}
\end{eqnarray}
\begin{eqnarray}\label{2DRicci}
\textrm{Non--vanishing Ricci tensor components}: \ \ \ \ \ \ R_{11}=-\frac{1}{4\sigma^2} \ , \ R_{22}=-\frac{1}{\sigma^2}
\end{eqnarray}
\begin{eqnarray}\label{2Dscalar}
\textrm{Scalar curvature}: \ \ \ \ \ \ R(r)=-\frac{1}{2}(1-r^2)       \ \ \ , \ \ \ -1\leq r\leq 1
\end{eqnarray}
From \eqref{2Dscalar} one can see that the curvature has a minimum value $R=-\frac{1}{2}$ when $r=0$ (absence of correlations) while for $|r|\rightarrow1$ (maximally correlated case) it has a maximum value $R=0$. In terms of the criterium of global chaos this can be interpreted as: the dynamics of the uncorrelated case is more chaotic than the corresponding to the maximally correlated case.
Moreover, the divergence of the metric observed for $|r|\rightarrow1$ expresses the maximally correlated case as a critical point of the dynamics.




\section{The ergodic hierarchy}\label{sec EH}
In classical chaos, the exponential instability implies continuous spectrum, and therefore, a decay of correlations in such a way that for large times the measure of the intersection between
two sets of phase space (separated from each other in time) tends to the product of their measures. This is the well known \emph{mixing} property, and constitutes one of the foundations of the statistical mechanics. The main feature of mixing is that it establishes the statistical independence of different parts of a trajectory, when sufficiently separated in time. This is the main reason for the application of probability theory
in the classical domain, which allows one to calculate statistical features such as diffusion, relaxation
and distribution functions \cite{casati libro}. Consequently, the description in terms of trajectories can be replaced by an equivalent one in terms of distribution functions, which, if not singular, represent not a single trajectory but a continuum of them.

In ergodic theory, any classical system is represented mathematically by a dynamical system $(X,\Sigma,\mu,\{T_t\}_{t\in J})$ where $X$ is a set, $\Sigma$ is a sigma--algebra of $X$, $\mu$ a measure defined over $\Sigma$ and $\{T_t\}_{t\in J}$ a group of measure--preserving transformations. The ergodic hierarchy ranks the chaos of a dynamical system according to a type of correlation $C(T_tA,B)$ between two subsets $A$ and $B$ of $X$ that are separated by a time $t$. This is defined as \cite{berko,lasota}
\begin{eqnarray}\label{3-1}
C(T_tA,B)=\mu(T_tA \cap B)-\mu(A) \ \mu(B)
\end{eqnarray}
The ergodic, mixing and Bernoulli levels of the EH are given in terms of (\ref{3-1}) in the following way. Given two arbitrary sets $A,B\in X$, it is said that $T_t$ is
\begin{enumerate}
  \item[$\bullet$] \emph{ergodic} if
\begin{eqnarray}\label{3-1ergodic}
\lim_{T\rightarrow\infty}\frac{1}{T}\int_{0}^{T}C(T_tA,B) \ dt=0 \ \ \ \ \ \ ,
\end{eqnarray}
  \item[$\bullet$] \emph{mixing} if
  \begin{eqnarray}\label{3-1mixing}
\lim_{t\rightarrow\infty}C(T_tA,B)=0 \ \ \ \ \ \ ,
\end{eqnarray}
\item[$\bullet$] \emph{Kolmogorov} if for all integer $r$, for all $A_0,A_1,\ldots,A_r \subseteq \Gamma$, and for all $\varepsilon>0$ there exists a positive integer $n_0>0$ such that if $n\geq n_0$ one has
\begin{eqnarray}\label{3-1kolmogorov}
|C(A_0,B)|<\varepsilon   \ \ \ \ \ \ , \ \ \ \ \ \textrm{for all} \  B\in \sigma_{n,r}(A_1,\ldots,A_r)
\end{eqnarray}
where $\sigma_{n,r}(A_1,\ldots,A_r)$ is the minimal $\sigma-$algebra generated by $\{T^k A_i:k\geq n \ ; \ i=1,...,r\}$.
  \item[$\bullet$] \emph{Bernoulli} if
  \begin{eqnarray}\label{3-1Bernoulli}
C(T_tA,B)=0 \ \ \ \ \ \ \ \  \textrm{for \ \  all} \ \ t \ \geq \ 0
\end{eqnarray}
  \end{enumerate}
In ergodic systems the correlation vanishes ``in time average" for large times while in mixing systems $C(T_tA,B)$ vanishes for $t\rightarrow\infty$. In Kolmogorov systems the correlations between an arbitrary set and another one belonging to the $\sigma$--algebra $\sigma_{n,r}(A_1,\ldots,A_r)$ cancel for $n\rightarrow\infty$. In Bernoulli systems the correlation is zero for all times. These levels classify the dynamics according to Eqs. (\ref{3-1ergodic}), (\ref{3-1mixing}), \eqref{3-1kolmogorov}, and (\ref{3-1Bernoulli}), from the weakest level (the ergodic) to the strongest (the Bernoulli). The following strict inclusions hold:
\begin{eqnarray}\label{2-6}
\textrm{ergodic} \ \supset \ \textrm{mixing} \ \supset \ \textrm{Kolmogorov} \ \supset \textrm{Bernoulli} \nonumber
\end{eqnarray}
In order to express $C(T_tA,B)$ by means of probability distributions it is more convenient to use the definition \eqref{3-1} in terms of distribution functions, which is given by \cite{lasota}
\begin{eqnarray}\label{3-2}
C(f\circ T_t,g)=\int_{X}(f\circ T_t)(x)g(x)dx-\int_{X}f(x)dx\int_{X}g(x)dx  \ \ \ \ \ \ \ \ \forall \ f,g \in \mathbb{L}^1(X)
\end{eqnarray}
where $f\circ T_t$ denotes the composition of $f$ and $T_t$, i.e. $f\circ T_t(x)=f(T_t(x))$ for all $x\in X$ and now the role of $A,B$ is played by the functions $f,g\in \mathbb{L}^1(X)$. Physically, $f$ represents any initial density function of the classical system whose value at time $t$ is given by $f\circ T_t$ with $T_t$ the classical Liouville evolution (in Hamiltonian systems).

\section{An information geometric version of the ergodic hierarchy}\label{sec IGEH}
Following the idea of characterizing chaos by means of the ergodic hierarchy \cite{paper1,paper2,paper3}, now we
consider an extension of the EH within the context of the information geometry.
In principle, in information geometry one has probability distributions $p_\theta$ that depend on a set of parameters $\theta$, and the dynamics of the macrovariables $\theta$ is performed along the geodesics of the statistical manifold.
Moreover, in the statistical manifold the role of time variable $t$ of dynamical systems is played by a parameter $\tau$ along the geodesics.

In order to introduce the tools of information geometry, we propose the following approach by defining a correlation between functions as the macrovariables $\theta$ evolve along the geodesics.
Given $N$ functions $f(x_i)$, each one of them in terms of the variable $x_i$ for all $i=1,\ldots,N$, we define the \emph{information geometric correlation} (IG correlation) $C(f_1,\ldots,f_N,\tau)$ between $f_1,\ldots,f_N$ at time--like parameter $\tau$ as
\begin{eqnarray}\label{3-4}
&C(f_1,\ldots,f_N,\tau)\doteq\nonumber\\
&\int p(x_1,\ldots,x_N;\theta(\tau))f_{1}(x_1)\cdots f_N(x_N)dx_1\cdots dx_N-\prod_{i=1}^{N}\int p_i(x_i;\theta(\tau))f_i(x_i)dx_i
\end{eqnarray}
where $\theta(\tau)=(\theta_1(\tau),\ldots,\theta_M(\tau))$ is the M--dimensional vector of the macrovariables at ``time" $\tau$ and,
\begin{eqnarray}\label{3-5}
p_i(x_i;\theta(\tau))=\int p(x_1,\ldots,x_N;\theta(\tau))\prod_{j\neq i}dx_{j} \ \ \ \ \ , \ \ \ \ i=1,\ldots,N
\end{eqnarray}
are the marginal distributions of $p(x_1,\ldots,x_N;\theta(\tau))$. From (\ref{3-4}) one can see that
$C(f_1,\ldots,f_N,\tau)$ measures how independent the variables $x_1,\ldots,x_N$ are at time--like $\tau$. This can be considered as a sort of information geometric generalization of the EH correlation.

Having established $C(f_1,\ldots,f_N,\tau)$ and taking into account the ergodic, mixing and Bernoulli levels given by Eqs. (\ref{3-1ergodic}), (\ref{3-1mixing}) and (\ref{3-1Bernoulli}), we define the \emph{information geometric ergodic hierarchy} (IGEH) as follows. For the sake of simplicity and since we focus mainly on the ergodicity and mixing properties (which are the fundamentals of statistical mechanics), in this contribution we do not extend the Kolmogorov level that involves a $\sigma$--algebra. Given a set of $N$ arbitrary functions $f_1(x_1),\ldots,f_N(x_N)$ we say that now the statistical model is
\begin{enumerate}
  \item[$\bullet$] \emph{IG ergodic} if
\begin{eqnarray}\label{3-5ergodic}
\lim_{T\rightarrow\infty}\frac{1}{T}\int_{0}^{T}C(f_1,\ldots,f_N,\tau)d\tau=0 \ \ \ \ \ \ ,
\end{eqnarray}
  \item[$\bullet$] \emph{IG mixing} if
  \begin{eqnarray}\label{3-5mixing}
\lim_{\tau\rightarrow\infty}C(f_1,\ldots,f_N,\tau)=0 \ \ \ \ \ \ ,
\end{eqnarray}
  \item[$\bullet$] \emph{IG Bernoulli} if
  \begin{eqnarray}\label{3-5Bernoulli}
C(f_1,\ldots,f_N,\tau)=0 \ \ \ \ \ \ \ \  \textrm{for \ \  all} \ \ t \ \in \ \mathbb{R}
\end{eqnarray}
  \end{enumerate}
As in the ergodic hierarchy, the following strict
inclusions hold:
\begin{eqnarray}
\textrm{IG ergodic}\ \supset \ \textrm{IG mixing} \supset \textrm{IG Bernoulli} \nonumber
\end{eqnarray}
For instance, a statistical model that is IG ergodic can be given by assuming that $C(f_1,\ldots,f_N,\tau)$ is proportional to $\sin(\alpha\tau)||f_1||_1\ldots||f_N||_1$ with $\alpha\in\mathbb{R}$. Making this replacement in \eqref{3-5ergodic} one obtains that $\lim_{T\rightarrow\infty}\frac{1}{T}\int_{0}^{T}C(f_1,\ldots,f_N,\tau)d\tau$ is equal to zero. Since $\sin(\alpha\tau)||f_1||_1\ldots||f_N||_1$ oscillates, then this model is not IG mixing nor IG Bernoulli. Examples of statistical models that are IG mixing and IG Bernoulli will be illustrated in Section \ref{sec relevance}.
Our approach, thus, deals with ergodic hierarchy in statistical models from an information geometry viewpoint.

\section{A measure of distinguishability for the $2D$ correlated model}\label{sec 2Dmeasure}
In order to use the levels of the IGEH for characterizing the dynamics of statistical models one should have a manner of determining the decay of the correlation $C(f_1,\ldots,f_N,\tau)$ in Eqs. (\ref{3-5ergodic}), (\ref{3-5mixing}) or (\ref{3-5Bernoulli}).
For the family of the $2D$ correlated probabilities $p(x,y|\mu_x,\sigma;r)$ of \eqref{2Dmodel}, we define a distinguishability measure $F:\{p(x,y;\mu,\sigma,r) \ \ | \ \ \mu_x\in(-\infty,\infty) \ , \ \sigma\in(0,\infty) \ , \ -1\leq r\leq1 \ \}\longmapsto \mathbb{R}$, given by
\begin{eqnarray}\label{4-1}
F(p)\doteq\|p(x,y;\mu_x,\sigma,r)-p_1(x)p_2(y)\|_{\infty}=\max_{(x,y)\in \mathbb{R}^2}\left|p(x,y;\mu_x,\sigma,r)-p_1(x)p_2(y)\right|
\end{eqnarray}
where $p_1(x),p_2(y)$ are the marginal distributions of $p(x,y;\mu_x,\sigma,r)$.
Furthermore, if $f_1(x),f_2(y)\in \mathbb{L}^1(\mathbb{R})$ are arbitrary functions of $x$ and $y$, then we have
\begin{eqnarray}\label{4-2}
&|C(f_1,f_2,\tau)|
=\left|\int_{\mathbb{R}^2}p(x,y;\mu_x,\sigma,r)f_1(x)f_2(y)dxdy-\int_{\mathbb{R}} p_1(x)f_1(x)dx\int_{\mathbb{R}} p_2(y)f_2(y)dy\right|\nonumber\\
&\leq \left\{\max_{(x,y)\in \mathbb{R}^2}\left|p(x,y;\mu_x,\sigma,r)-p_1(x)p_2(y)\right|\right\} \left|\int_{\mathbb{R}^2} dxdy f_1(x)f_2(y)\right| = F(p)||f_1f_2||_{1}
\end{eqnarray}
Eq. \eqref{4-2} expresses that $F(p)||f_1f_2||_{1}$ is un upper bound for $|C(f_1,f_2,\tau)|$. Therefore, it is convenient to find an analytic expression for \eqref{4-1}. After some algebra one can obtain that\footnote{The demonstration can be found in the Appendix.}
\begin{eqnarray}\label{4-3}
F(p)=|r|\left(\sqrt{1-r^2}(1+|r|)\right)^{-1-\frac{1}{|r|}} \ \ \ \ \ \ \textrm{for \ all} \  r \ \in \ [-1,1]
\end{eqnarray}
\begin{figure}[!!ht]
\begin{center}
\includegraphics[width=12cm]{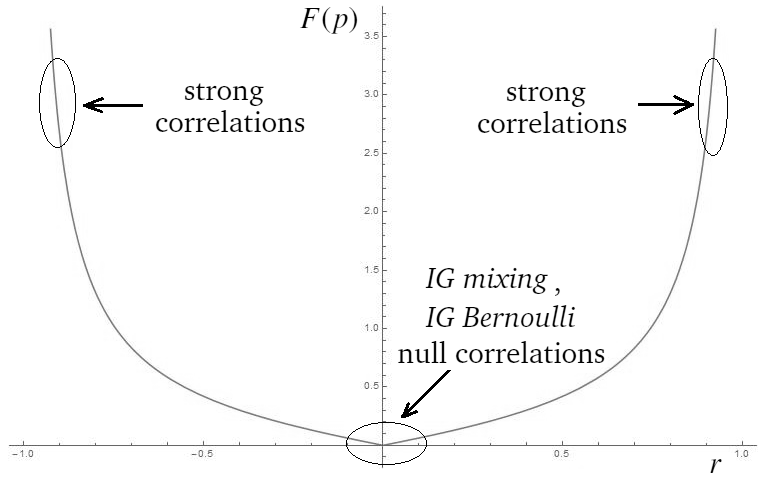}
\end{center}
  \caption{Behavior of $F(p)$ in terms of the correlation coefficient. Near to the region $r=0$ the statistical model belongs to the IG mixing level while for $r=0$ it is IG Bernoulli. When $r\rightarrow\pm1$ one has that $F(p)$ diverges with the presence of strong correlations. The discontinuity of the slope of $F$ at $r=0$ is due to the particular form of $F(p)$, i.e. the maximum operation of infinite norm can present discontinuities in its derivatives.}\label{fig:distinguible}\end{figure}
The behavior of $F(p)$, which is independent of $\mu_x$ and $\sigma$, is shown in Fig. 1. Two relevant regions, corresponding to the limiting cases $r\rightarrow0$ and $r\rightarrow\pm1$, can be well distinguished. The region $r\rightarrow0$ corresponds to the zone where the statistical model is characterized by the IG mixing and IG Bernoulli levels, with the particularity that the variables of microspace are uncorrelated. Moreover, one can see that near to $r=0$ the decay is linear in $r$. The curve $F(p)$ also shows that, if $r\rightarrow0$ when $\tau\rightarrow\infty$, then the statistical model is IG mixing.

In the region $r\rightarrow\pm1$ the measure $F(p)$ diverges corresponding to the maximally correlated case, which physically means that the system presents strong correlations between the variables of microspace. Due to the correlations are strong in this regime the statistical model cannot be IG mixing nor IG Bernoulli.

Finally, it should be noted that $F(p)$ does not allow one to distinguish between two probability distributions having $r$ and $-r$ respectively. The symmetry respect to the axis $r=0$ is due to the mathematical form of the infinite norm $||.||_{\infty}$ in the definition (\ref{4-1}). That is, with other choices of $F(p)$ one could distinguish states (probability distributions) with correlation coefficients $r$ and $-r$.

\section{Geometrizing the canonical ensemble by means of statistical models and the IGEH}\label{sec geometrical canonical}

On the basis of the above characterization of the dynamics of the macrospace in terms the IGEH, our next aim is to give a method for studying the dynamics of a system belonging to the canonical ensemble.

\subsection{Canonical ensembles in the context of the information geometry}\label{subsec canonical}
Beyond the relationship between statistical models and statistical physics has been already established \cite{amari,expo1}, extensions in several directions have been recently introduced \cite{expo2,expo3,expo4,expo5,expo6,expo7,expo8,expo9}, with a particular focusing on the exponential families since they represent mathematically the Liouville densities of the statistical ensembles. Relevant consequences from these researches such as the connection between Hessian structures and exponential families \cite{expo8}, nonextensive statistical models \cite{expo4,expo5,expo6,expo9} and other extensions \cite{expo2,expo3,expo7} has been reported.

In order to apply the IGEH to the canonical ensemble, here we obtain some explicit formulas that relate the statistical parameters of multivariate Gaussian distributions (which are a special case of exponential family) with the physical parameters of quadratic Hamiltonians. In the present contribution we only focus on the multivariate Gaussian distributions and we will omit the definition of exponential families of a more general character.
We begin by considering the family of probability density functions given by the classical canonical ensemble
\begin{eqnarray}\label{canonical1}
p(q,p;\theta)=\frac{1}{Z(\theta)}\exp\left\{-\beta E(q,p,\theta)\right\}
\end{eqnarray}
where $Z(\theta)=\int p(q,p;\theta)dqdp$ is the well known partition function, $\beta=\frac{1}{k_B T}$ is the Boltzmann factor, $E(q,p,\theta)$ is the energy of the system expressed in terms of the phase space coordinates $(q,p)\in\Gamma$ (with $\Gamma$ the phase space), and $\theta$ are the macrovariables of the system.
The function $p(q,p;\theta)$ represents the probability density of the microstate $(q,p)$ of the system, corresponding to the macrostate given by the macroscopic parameters $\theta$, when contacted in thermal equilibrium with a heat bath at a fixed temperature $T$.
In this way, one can see that there is a biunivocal correspondence between statistical models and statistical ensembles: each pair of microvariables $(q,p)$ corresponds to a microstate of the ensemble, and each value of the macrovariable $\theta$ is associated to a macrostate of the system.

Assuming a $2n$--dimensional phase space $\Gamma$ and a $m$--dimensional macrospace $\Theta$, from \eqref{canonical1} and using \eqref{fisherao} one obtains the Fisher tensor for the canonical ensemble
\begin{eqnarray}\label{canonical2}
&g_{ij}(\theta)_{CE}=\frac{\beta^2}{Z(\theta)}\int dqdp \exp\left\{-\beta E(q,p,\theta)\right\}\frac{\partial E(q,p;\theta)}{\partial \theta_i}\frac{\partial E(q,p;\theta)}{\partial \theta_j}\nonumber\\
&=\beta^2 \left\langle \frac{\partial E(q,p;\theta)}{\partial \theta_i}\frac{\partial E(q,p;\theta)}{\partial \theta_j}\right\rangle_{CE}  \ \ \ \ \ \ \ \ \ \ \ \ \ \ \ i,j=1,\ldots,m
\end{eqnarray}
where $q$ and $p$ are a short notation for $(q_1,\ldots,q_n)$ and $(p_1,\ldots,p_n)$, and the subffix $CE$ stands for the canonical ensemble. The formula \eqref{canonical2} is the expression of the canonical ensemble in the context of the information geometry. All the dynamics over the macrospace can be derived by means of the Eqs. \eqref{geodesic}--\eqref{scalar}, thus obtaining a geometrical characterization of the canonical ensemble.

\subsection{Dynamics of the canonical ensemble in terms of the IGEH levels}\label{subsec dynamics IGEH}

The particular dependence of the energy $E(q,p;\theta)$ on the microvariables $(q,p)$ determines the correlations between them, which are reflected in the probability distribution $p(q,p;\theta)$.
Since the IG correlation $C(f_1,\ldots,f_N,\tau)$ measures the degree of statistical independence between the microvariables of the microspace at time--like parameter $\tau$, then for a given expression of the energy it is desirable to know what is the form of $C(f_1,\ldots,f_N,\tau)$.
We focus on a particular form of the energy, i.e. when $E(q,p;\theta)$ is a quadratic function of $(q,p)$. The physical relevance of this assumption lies, among other things, in the fact that it allows to study the dynamics of the system near of an equilibrium point. In this case, the formula of the energy is given by
\begin{eqnarray}\label{canonical3}
&E(q,p;\theta)=(q-q_0,p-p_0) \ \textbf{G} \ (q-q_0,p-p_0)^{T} \nonumber\\
& \nonumber\\
& \textbf{G}_{ij}=\frac{1}{2}\frac{\partial^2E}{\partial x_i \partial x_j}(q_0,p_0) \ \ \ \ \ , \ \ \ \ i,j=1,\ldots,n \\
&\nonumber\\
&(x_1,\ldots,x_n,x_{n+1},\ldots,x_{2n})=(q_1,\ldots,q_n,p_{1},\ldots,p_{n})\nonumber
\end{eqnarray}
with $\textbf{G}$ the Hessian of $E$ around at a point $(q_0,p_0)$ and $(q-q_0,p-p_0)^{T}$ the transposed of $(q-q_0,p-p_0)$. Then, the probability distribution $p(q,p;\theta)$ adopts the form
\begin{eqnarray}\label{canonical4}
p(q,p;\theta)=\frac{1}{Z(\theta)}\exp\left\{-\beta (q-q_0,p-p_0) \ \textbf{G} \ (q-q_0,p-p_0)^{T}\right\}
\end{eqnarray}
with $Z(\theta)^{-1}$ the normalization factor. From \eqref{canonical4} one can see that $p(q,p;\theta)$ is nothing but a $2n$--multivariate Gaussian distribution on the microvariables $q_1,\ldots,q_n,p_{1},\ldots,p_{n}$. At this point it is convenient to recall the expression of the $2n$--multivariate Gaussian distribution $p(x;\mu,\boldsymbol{\Sigma})$
\begin{eqnarray}\label{canonical5}
p(x;\mu,\boldsymbol{\Sigma})=\left(\frac{1}{2\pi}\right)^n \frac{1}{\sqrt{|\boldsymbol{\Sigma}|}}\exp\left\{-\frac{1}{2} (x-\mu) \ \boldsymbol{\Sigma}^{-1} \ (x-\mu)^{T}\right\}
\end{eqnarray}
with $x=(x_1,\ldots,x_{2n})$ the vector of microvariables, $\mu=(\mu_1,\ldots,\mu_{2n})$ the mean value vector, $\boldsymbol{\Sigma}$ the covariance matrix and $\boldsymbol{\Sigma}$ its inverse, and $|\boldsymbol{\Sigma}|$ the determinant of $\boldsymbol{\Sigma}$. Moreover, the marginals
$p_i(x_i;\mu_i,\boldsymbol{\Sigma}_{ii})$
of $p(x;\mu,\boldsymbol{\Sigma})$ are given by
\begin{eqnarray}\label{canonical6}
p_i(x_i;\mu_i,\boldsymbol{\Sigma}_{ii})=\frac{1}{\sqrt{2\pi \boldsymbol{\Sigma}_{ii}}}\exp\left\{-\frac{1}{2} \frac{1}{\boldsymbol{\Sigma}_{ii}}(x_i-\mu_i)^2\right\} \ \ \ \ \ , \ \ \ \ i=1,\ldots,2n
\end{eqnarray}
where $\boldsymbol{\Sigma}_{ij}$ stands for the $ij$--th matrix element of $\boldsymbol{\Sigma}$ with $i,j=1,\ldots,2n$.

Let us consider $2n$ arbitrary functions $f_1(x_1),\ldots,f_{2n}(x_n)\in\mathbb{L}^1(\mathbb{R})$. Then, replacing \eqref{canonical5} and \eqref{canonical6} in \eqref{3-4} one has that
\begin{eqnarray}\label{canonical7}
&C(f_1,\ldots,f_{2n},\tau)\doteq\nonumber\\
&\int_{\mathbb{R}^{2n}} \left[\left(\frac{1}{2\pi}\right)^n \frac{1}{\sqrt{|\boldsymbol{\Sigma}|}}\exp\left\{-\frac{1}{2} (x-\mu) \ \boldsymbol{\Sigma}^{-1} \ (x-\mu)^{T}\right\}f_1(x_1)\cdots f_{2n}(x_{2n}) \right]dx_1\cdots dx_{2n}\\
&- \ \prod_{i=1}^{2n}\int_{\mathbb{R}}\left[\frac{1}{\sqrt{2\pi \boldsymbol{\Sigma}_{ii}}}\exp\left\{-\frac{1}{2} \frac{1}{\boldsymbol{\Sigma}_{ii}}(x_i-\mu_i)^2\right\}f_i(x_i)\right]dx_i \nonumber
\end{eqnarray}
is the IG correlation for the family of $2n$--multivariate Gaussian distributions. As in the $2D$ correlated model, one can obtain an upper bound
for $C(f_1,\ldots,f_{2n},\tau)$ as follows.
\begin{eqnarray}\label{canonical8}
&|C(f_1,\ldots,f_{2n},\tau)|\leq \nonumber\\
&\max_{x\in \mathbb{R}^{2n}}\left\{\left(\frac{1}{2\pi}\right)^n \frac{1}{\sqrt{|\boldsymbol{\Sigma}|}}\exp\left\{-\frac{1}{2} (x-\mu) \ \boldsymbol{\Sigma}^{-1} \ (x-\mu)^{T}\right\} - \prod_{i=1}^{2n}\frac{1}{\sqrt{2\pi \boldsymbol{\Sigma}_{ii}}}\exp\left\{-\frac{1}{2} \frac{1}{\boldsymbol{\Sigma}_{ii}}(x_i-\mu_i)^2\right\} \right\} \nonumber \\
& \times \ \left|\int_{\mathbb{R}^{2n}} f_1(x_1)\cdots f_{2n}(x_{2n})dx_1\cdots dx_{2n} \right|=\\
&\max_{x\in \mathbb{R}^{2n}}\left\{\left(\frac{1}{2\pi}\right)^n \frac{1}{\sqrt{|\boldsymbol{\Sigma}|}}\exp\left\{-\frac{1}{2} (x-\mu) \ \boldsymbol{\Sigma}^{-1} \ (x-\mu)^{T}\right\} - \prod_{i=1}^{2n}\frac{1}{\sqrt{2\pi \boldsymbol{\Sigma}_{ii}}}\exp\left\{-\frac{1}{2} \frac{1}{\boldsymbol{\Sigma}_{ii}}(x_i-\mu_i)^2\right\} \right\}||f_1(x_1)\cdots f_{2n}(x_{2n})||_1 \nonumber
\end{eqnarray}
This inequality expresses an upper bound of the IG correlation for the family of the $2n$--multivariate Gaussian distributions, where the maximum is a measure of distinguishability, and the parameters $\mu=\mu(\tau)$, $\boldsymbol{\Sigma}=\boldsymbol{\Sigma}(\tau)$ are dependent on $\tau$ along the geodesics by means of the application of Eqs. \eqref{fisherao}--\eqref{christoffel} to $p(x;\mu,\boldsymbol{\Sigma})$.

Now we can set the upper bound of \eqref{canonical8} in the language of the canonical ensemble. By simple inspection of Eqs. \eqref{canonical3}--\eqref{canonical6}, if one makes the following replacements
\begin{eqnarray}\label{canonical9}
&x=(x_1,\ldots,x_{2n}) \ \longrightarrow \ (q,p)=(q_{1},\ldots,q_{n},p_{1},\ldots p_{n})\nonumber\\
&\nonumber\\
&\mu=(\mu_1,\ldots,\mu_{2n}) \ \longrightarrow \ (q_0,p_0)=(q_{10},\ldots,q_{n0},p_{10},\ldots p_{n0})\nonumber\\
&\\
&|\boldsymbol{\Sigma}|^{-1}\ \longrightarrow \ |2\beta \textbf{G}_{ij}|=(2\beta)^{2n}|\textbf{G}_{ij}| \nonumber\\
&\nonumber\\
&\boldsymbol{\Sigma}^{-1}_{ij} \ \longrightarrow \ 2\beta \textbf{G}_{ij}=\beta\frac{\partial^2E}{\partial x_i \partial x_j}(q_0,p_0) \ \ \ \ \ \ \ \ , \ \ \ \ \ \ \ \ i,j=1,\ldots,2n\nonumber
\end{eqnarray}
in \eqref{canonical8} then one obtains
\begin{eqnarray}\label{canonical10}
&|C(f_1,\ldots,f_{2n},\tau)_{CE}|\leq \nonumber\\
&\max_{(q,p)\in \Gamma}[\left(\frac{\beta}{\pi}\right)^n \frac{1}{\sqrt{|\textbf{G}^{-1}|}}\exp\left\{-\beta (q-q_0,p-p_0) \ \textbf{G} \ (q-q_0,p-p_0)^{T}\right\}\nonumber\\
& - \prod_{i=1}^{n}\frac{\beta}{\pi\sqrt{(\textbf{G}^{-1})_{ii}(\textbf{G}^{-1})_{n+in+i}}}\exp\left\{-\beta  \left(\frac{1}{(\textbf{G}^{-1})_{ii}}(q_i-q_{i0})^2+\frac{1}{(\textbf{G}^{-1})_{n+in+i}}(p_i-p_{i0})^2\right)\right\}]\\
&\times \ ||f_1(q_1)\cdots f_n(q_n)f_{n+1}(p_1)\cdots f_{2n}(p_{n})||_1 \nonumber
\end{eqnarray}
The inequality \eqref{canonical10} is an upper bound of the IG correlation for a system in the canonical ensemble whose energy is a quadratic form, expanded around a point $(q_0,p_0)$ of the phase space. Since the macrovariables
$(q_0,p_0)$ and $\textbf{G}$ are functions of the time--like parameter $\tau$ then the utility of \eqref{canonical10} is that, according to the way the IG correlation cancel for large values of $\tau$, one can classify the statistical model as belonging to some of the levels of the IGEH. In this way, one can study phase transitions in terms of the IGEH levels as the macrovariables vary. For instance, if the dynamics in macrospace is such that the maximum in \eqref{canonical10} tends to zero when $\tau\rightarrow\infty$, then one has that the canonical ensemble behaves as an IG mixing statistical model.

It should be noted that the multivariate Gaussian distribution only exists if the covariance matrix $\boldsymbol{\Sigma}$ is positive, which implies that $\boldsymbol{\Sigma}^{-1}$ must be also positive. Then, it follows that $\frac{\partial^2 E}{\partial^2 x_i}(q_0,p_0)>0$. In particular, the stable equilibrium points of the system satisfy this condition.

Now we are in a position to reach one of our main contributions of this work. By the Eq. \eqref{canonical9} and using the definition of the tensor $\textbf{G}$, one finally obtains
\begin{eqnarray}\label{canonical12}
|\boldsymbol{\Sigma}|\left|\frac{\partial^2E}{\partial x_i \partial x_j}(q_0,p_0)\right|  = (k_B T)^{2n}
\end{eqnarray}
This equation expresses an intimate relationship between the canonical ensemble and the statistical models, which one can express in words as follows:

\emph{``given a quadratic form of the energy, the determinant of the covariance matrix which measures the correlations between the microvariables, is proportional to a power (equal to the dimension of the phase space) of the temperature of the heat bath"}.

Moreover, since the determinant of covariance matrix is a decreasing function of the correlations, as the temperature of the heat bath increases the correlations tend to be suppressed, as expected statistically. This result will be useful for characterizing phase transitions, as we shall see below. 

\section{Models and results}\label{sec relevance}
In order to illustrate the relevance of the IGEH we consider two examples belonging to different topics: an interacting bipartite system presenting phase transitions in the canonical ensemble, and the $2\times2$ Gaussian orthogonal ensemble. Next, we give a panoramic outlook of the IGEH and of the relationship between statistical models and Liouville densities of the canonical ensemble.

\subsection{Phase transitions in a pair of interacting harmonic oscillators}\label{subsec harmonic oscillators}

Let us consider a pair of unidimensional and interacting harmonic oscillators in the canonical ensemble whose total energy is given by
\begin{eqnarray}\label{app1}
&E(q_1,q_2,p_1,p_2;q_{10},q_{20},m_1,m_2,\omega_1,\omega_2,r)=T(p_1,p_2;m_1,m_2)+V(q_1,q_2;q_{10},q_{20},m_1,m_2,\omega_1,\omega_2,r)\nonumber\\
&\nonumber\\
&T(p_1,p_2;m_1,m_2)=\frac{p_1^2}{2m_1}+\frac{p_2^2}{2m_2} \nonumber\\
&\nonumber\\
&V(q_1,q_2;q_{10},q_{20},m_1,m_2,\omega_1,\omega_2,r)=\frac{1}{2}m_1\omega_1^2(q_1-q_{10})^2+\frac{1}{2}m_2\omega_2^2(q_2-q_{20})^2-r\sqrt{m_1m_2}\omega_1\omega_2(q_1-q_{10})(q_2-q_{20})\nonumber
\end{eqnarray}
where $q_i$, $p_i$, $q_{i0}$, $m_i$, and $\omega_i$ are the position, the momentum, the equilibrium position, the mass, and the frequency of the $i$--th particle with $i=1,2$. Here $T(p_1,p_2;m_1,m_2)$ is the kinetic energy and $V(q_{10},q_{20},m_1,m_2,\omega_1,\omega_2,r)$ is the potential energy which is composed by three terms: the first two are the potential energy of each oscillator separatelly while the term $-r\sqrt{m_1m_2}\omega_1\omega_2(q_1-q_{10})(q_2-q_{20})$ represents the interaction between the oscillators, and the coefficient $r\in[-1,1]$ measures the strength coupling. For instance, when $r=0$ the oscillators are uncoupled, and therefore, their motions are independent of each other.

Now, in order to use the analysis made about the $2D$ correlated model one must reduce the number of microvariables and macrovariables, and also impose some type of constraints. In this sense, we fix the masses $m_1,m_2$ and set $q_{20}=0$. Also, we consider the following constraint
\begin{eqnarray}\label{app2}
\Sigma^2=\frac{k_BT_0}{\sqrt{m_1m_2}\omega_1\omega_2}
\end{eqnarray}
where $\Sigma$ is a real constant, $k_B$ is the Boltzmann constant, and $T_0$ is a temperature of reference\footnote{For instance, the room temperature $\sim 20^{\circ}$ C (293.15 K).}. Here $T_0$ plays the role of being a temperature that breaks up the correlations between the microvariables.

Due to the correlations are only between the position coordinates $q_1,q_2$ then one can reasonably neglect the momentum coordinates $p_1,p_2$ by integrating the probability distribution $p(q_1,q_2,p_1,p_2;\theta)$ given by the canonical ensemble, i.e.
$\frac{1}{Z(\theta)}\exp\{-\beta E(q_1,q_2,p_1,p_2;\theta)\}$, over $p_1$ and $p_2$. And since only the kinetic energy has the dependence on $p_1$ and $p_2$ then this equivalent to consider a sort of marginal probability distribution with respect to the potential energy, i.e.
\begin{eqnarray}\label{app3}
p(q_1,q_2;\theta)=A(\theta)\exp\{-\beta V(q_1,q_2;\theta)\}   \ \ \ \ \ \  \textrm{with} \ \ \ \ \ \ A(\theta)^{-1}=\int \exp\{-\beta V(q_1,q_2;\theta)\}dq_1dq_2
\end{eqnarray}
With the help of \eqref{app2} one can express the potential energy as
\begin{eqnarray}\label{app4}
V(q_1,q_2;q_{10},\omega_1,r)=\frac{k_B T_0}{2}\left\{\frac{(q_1-q_{10})^2}{k_B T_0(m_1\omega_1^2)^{-1}}+\frac{q_2^2 k_B T_0(m_1\omega_1^2)^{-1}}{\Sigma^4}-\frac{2r(q_1-q_{10})q_2}{\Sigma^2}\right\}
\end{eqnarray}
Then, from \eqref{app3} and \eqref{app4} one has
\begin{eqnarray}\label{app5}
&p\left(q_1,q_2;q_{10},\sqrt{k_B T_0(m_1\omega_1^2)^{-1}},r\right)=A\left(q_{10},\sqrt{k_B T_0(m_1\omega_1^2)^{-1}},r\right)\nonumber\\
&\times \ \exp\left\{-\beta\frac{k_B T_0}{2}\left\{\frac{(q_1-q_{10})^2}{k_B T_0(m_1\omega_1^2)^{-1}}+\frac{q_2^2k_B T_0(m_1\omega_1^2)^{-1}}{\Sigma^4}-\frac{2r(q_1-q_{10})q_2}{\Sigma^2}\right\}\right\}
\end{eqnarray}
which is nothing but the probability distribution of the $2D$ correlated model \eqref{2Dmodel} by means of the identifications
\begin{eqnarray}\label{app6}
&(x,y) \ \longleftrightarrow \ (q_1,q_2) \nonumber\\
& \nonumber\\
&\mu_x \ \longleftrightarrow \ q_{10} \nonumber\\
& \\
&\sigma \ \longleftrightarrow \ \sqrt{k_B T_0(m_1\omega_1^2)^{-1}} \nonumber\\
& \nonumber\\
& (1-r^2)^{-1} \ \longleftrightarrow \ \beta k_B T_0=\frac{T_0}{T} \nonumber
\end{eqnarray}
From the last line of \eqref{app6} one obtains
\begin{eqnarray}\label{app7}
1-r^2=\frac{T}{T_0}
\end{eqnarray}
Let us show that this equation is a particular case of the formula \eqref{canonical12}: since the determinant of the covariance matrix is $\Sigma^4(1-r^2)$ and the determinant of the Hessian of the potential energy evaluated at $(q_{10},q_{20})$ is \\
$(1-r^2)m_1m_2\omega_1^2\omega_2^2$ then one can replace both expressions in \eqref{canonical12} with $n=1$, thus obtaining
\begin{eqnarray}\label{app9}
\Sigma^4(1-r^2)^2m_1m_2\omega_1^2\omega_2^2=(k_BT)^2 \nonumber
\end{eqnarray}
With the help of \eqref{app2} one can recast this equation as
\begin{eqnarray}\label{app10}
\Sigma^4(1-r^2)^2(k_BT_0)^2\frac{1}{\Sigma^4}=(k_BT)^2 \nonumber
\end{eqnarray}
from which one obtains $1-r^2=\frac{T}{T_0}$.

Considering the temperature $T$ as an external parameter one can study the phase transitions of the system as $T$ varies. Also, we set the reference temperature $T_0$ as the room temperature. Given that the parameter $\tau$ is arbitrary, we choose $\tau$ as
\begin{eqnarray}\label{app11}
\tau=\frac{1}{1-\frac{T}{T_0}} \nonumber
\end{eqnarray}
This choice for $\tau$ is convenient since one has
\begin{eqnarray}\label{app12}
&\tau\longrightarrow\infty \ \ \ \ \ \ \textrm{if and only if} \ \ \ \ \ \ T\longrightarrow T_0 \nonumber\\
\end{eqnarray}
In this way, the asymptotic limit $\tau\rightarrow\infty$ is identified with the limit $T\rightarrow T_0$, and therefore, the transition towards high temperatures can be studied by means of the limit $\tau\rightarrow\infty$. This transition express the behavior of the correlations between the oscillators when the bath temperature pass from a finite value (which is lower than $T_0$) to the room temperature. Physically, at the room temperature is expected that if the energy $k_BT_0$ delivered by the bath to each oscillator is larger than the energies of them (i.e., $k_BT_0\gg m_{1,2}\omega_{1,2}^2$) then as a result of the thermal agitation the correlations between the oscillators tend to be canceled. Indeed, from \eqref{app7} one can see that $r\rightarrow0$ when $T\rightarrow T_0$.

Now let us see that this phase transition is characterized in terms of the mixing level of the IGEH. When $r$ is vanishingly small, the following approximations hold:
\begin{eqnarray}\label{app13}
&(\sqrt{1-r^2})^{-1-\frac{1}{|r|}}\approx 1-r^2(-\frac{1}{2}-\frac{1}{2|r|})=1+\frac{1}{2}|r|+\frac{1}{2}r^2 \nonumber\\
&(1+|r|)^{-1-\frac{1}{|r|}}\leq 1 \nonumber
\end{eqnarray}
Using these approximations, and neglecting terms of order $r^2$, in the formula \eqref{4-3} of $F(p)$ one obtains that $F(p)\lesssim |r|$ holds for $|r|\ll1$. Replacing this inequality in \eqref{4-2} one has that
\begin{eqnarray}\label{app13}
|C(f_1,f_2,\tau)|\lesssim |r|||f_1f_2||_1    \ \ \ \ \ , \ \ \ \ \ \textrm{for} \ \ |r|\ll1 \nonumber
\end{eqnarray}
In turn, since $r\rightarrow0$ when $\tau\rightarrow\infty$ this equation implies
\begin{eqnarray}\label{app13}
\lim_{\tau\rightarrow\infty}|C(f_1,f_2,\tau)|\lesssim \lim_{\tau\rightarrow\infty}|r|||f_1f_2||_1=0
\end{eqnarray}
According to the definition \eqref{3-5mixing} then the system is \emph{IG mixing}. This is the regime of null correlations corresponding to the region of the curve of $F(p)$ around at $r=0$, as can be seen in Fig. 1. When $r=0$ the probability distribution \eqref{app5} can be factorized as the product of its marginals, and therefore, from the Eq. \eqref{3-5Bernoulli} one has that the system is \emph{IG Bernoulli}.

Moreover, since the scalar curvature for the $2D$ correlated model is $R=-\frac{1}{2}(1-r^2)$ then one obtains
\begin{eqnarray}\label{app14}
R=-\frac{T}{2T_0}
\end{eqnarray}
The formula \eqref{app14} expresses the connection between the thermodynamics of the canonical ensemble and the information geometry of the system of coupled oscillators. It can be seen that the statistical model behaves as an ``intermediary" between the thermodynamic parameters and the geometrical quantities of the statistical manifold. As a consequence, the determinant of the covariance matrix of the statistical model is determined by the Boltzmann factor, thus linking the temperature with all the geometrical quantities like the metric tensor, the scalar curvature, etc.

From Eq. \eqref{app14} it follows that the scalar curvature decreases as the bath temperature increases up to reach a minimum value $R=-\frac{1}{2}$ at the room temperature, where the correlation coefficient is zero. On the other hand, when the temperature tends to zero the scalar curvature takes a maximum value $R=0$ which corresponds to the maximally correlated case $r\rightarrow\pm1$. This reflects the intuitively image that in absence of thermal agitation the correlations remain present.

Finally, it should be noted that this analysis is consistent with the Cafaro's criteria of global chaos: as the temperature grows the dynamics become more chaotic and the scalar curvature turns out more negative. From the point of view of the IGEH this is characterized in terms of the mixing level, in which the breaking up of the coupling between the oscillators at the room temperature is expressed by means of the cancellation of the IG correlation in the asymptotic limit.



\subsection{$2\times2$ Gaussian Orthogonal Ensembles (GOE)}\label{subsec Gaussian ensembles}

In Gaussian Orthogonal Ensembles theory one deals with the probability distribution $p(H_{11},H_{12},\ldots,H_{nn})$ for the Hamiltonian matrix elements assuming that the $H_{ij}$ are uncorrelated \cite{stockmann,haake}. Then in the framework of information geometry one could try to describe them by defining
a microspace $x_1,x_2,\ldots,x_n$ and a macrospace $\theta_1,\theta_2,\ldots,\theta_m$ in a suitable way.

In order to characterize the GOE within a statistical model we study a correlated ensemble of $2\times 2$ matrices.
We take the microspace as the Hamiltonian matrix elements $\{H_{11},H_{22},H_{12},H_{21}\}$ and define the macrospace as follows. 
For the sake of simplicity, we choose the macrospace in such way that only $H_{11}$ and $H_{22}$ are correlated, and that the mean values of all variables are zero, except for the mean value corresponding to $H_{11}$ which is equal to $\mu$. Also, we consider that the variance of $H_{11}$, $H_{12}$ and $H_{21}$ are the same, denoted by $\sigma$. Moreover, in order to study how independent the diagonal Hamiltonian elements are, we restrict the dynamics by considering that $r\in[-1,1]$ is the correlation coefficient between $H_{11}$ and $H_{22}$, and that the product of the covariances between $H_{11}$ and $H_{22}$ is a constant $\Sigma^2$.
Taking this into account, the resulting macrospace is $\{(\mu,\sigma)\in \mathbb{R}\times(0,\infty)\}$ and the correlated probability distribution is given by \footnote{Note that, since the GOE correspond to the orthogonal class of Hamiltonians then one has that $H_{12}=H_{21}$. However, in the formalism of Random matrices and for the orthogonal case, the volume element $dH_{11}dH_{22}dH_{12}dH_{21}$ (as if $H_{12}$ and $H_{21}$ were independent variables) is the real Lebesgue measure of $\mathbb{R}^4$ and must to be taken into account in order to normalize the probability distribution \cite{stockmann}.}
\begin{eqnarray}\label{GOE joint}
&p(H_{11},H_{22},H_{12},H_{21};\mu,\sigma,r)=\nonumber\\
&\frac{1}{2\pi\Sigma^2\sqrt{1-r^2}}\exp\left(-\frac{1}{2(1-r^2)}\left[\frac{(H_{11}-\mu)^2}{\sigma^2}+\frac{H_{22}^2\sigma^2}{\Sigma^4}-
\frac{2r(H_{11}-\mu)H_{22})}{\Sigma^2}\right]\right)
\frac{1}{2\pi\sigma^2}\exp\left(-\frac{(H_{12}^2+H_{21}^2)}{2\sigma^2}\right) \nonumber
\end{eqnarray}
where the correlation coefficient $r$ is considered as an external parameter and since the correlation between $H_{11}$ and $H_{22}$ is in terms of $r$ then $\Sigma$ can be taken as a fixed constant. In turn, given that $H_{11}$ and $H_{22}$ are the only microvariables correlated to each other and since the transitions of the dynamics depend fundamentally on the correlations, then one can reasonably neglect $H_{12}$ and $H_{21}$. Analogously as it was made in the pair of oscillators, one can integrate the correlated probability distribution over $H_{12}$ and $H_{21}$, thus obtaining
\begin{eqnarray}\label{GOE marginal}
p(H_{11},H_{22};\mu,\sigma,r)=\frac{1}{2\pi\Sigma^2\sqrt{1-r^2}}\exp\left(-\frac{1}{2(1-r^2)}\left[\frac{(H_{11}-\mu)^2}{\sigma^2}+\frac{H_{22}^2\sigma^2}{\Sigma^4}-
\frac{2r(H_{11}-\mu)H_{22})}{\Sigma^2}\right]\right)
\end{eqnarray}
which is nothing but the $2D$ correlated model, i.e.  \eqref{2Dmodel} and \eqref{GOE marginal} identical by renaming $H_{11}$ and $H_{22}$ as $x$ and $y$ respectively.
Then, the non vanishing components of the Ricci tensor $R_{ij}$ and the Ricci scalar curvature $R$ are given by the Eqs. \eqref{2DRicci} and \eqref{2Dscalar}
\begin{eqnarray}\label{GOE Ricci tensor}
R=g^{11}R_{11}+g^{22}R_{22}=-\frac{1}{2}(1-r^2) \ \ \ \ \ \ \  ,\textrm{with} \ \ \ \ \ \ \  R_{11}=-\frac{1}{4\sigma^2} \ , \ R_{22}=-\frac{1}{\sigma^2} \nonumber
\end{eqnarray}
Three remarks follow. First, the statistical manifold has a curvature which is negative
for all values of the correlation coefficient $r\in[-1,1]$. Based on the Cafaro's criterium above, this simply means that the dynamics in macrospace $(\mu,\sigma)$ is chaotic for all $r$.

Second, the $2\times 2$ GOE case corresponds to $r=0$ and $\Sigma=\sigma$, thus having the minimum value of the scalar curvature
\begin{eqnarray}\label{GOE chaotic}
R_{GOE}=R(r=0)=-\frac{1}{2}=R_{\min} \ \ \ \ \ \ \ \ \ \ \ \ \ \ \ \ \ \ \ \ \ \ (\textrm{GOE, most chaotic case}) \nonumber
\end{eqnarray}
In this case the correlated probability distribution is the product of their marginals and thus, the model is \emph{IG Bernoulli}. Therefore, one can see that the GOE corresponds to the strongest level of the IGEH and this can be considered as a characterization of the Gaussian ensembles from an information geometric point of view.

Third, for the strongly correlated case that corresponds to $|r|\sim1$ one has
\begin{eqnarray}\label{GOE correlated}
R(|r|\rightarrow1)=0 \ \ \ \ \ \ \ \ \ \ \ \ \ \ \ \ \ \ \ \ \ \ (\textrm{strongly correlated case}) \nonumber
\end{eqnarray}
which can be interpreted, by the Cafaro's criterium of global chaos, as the case when the dynamics is the least chaotic of all.

\subsection{A panoramic outlook of the IGEH and of the canonical ensemble in curved statistical models}\label{subsec outlook IGEH}
\begin{figure}[!!ht]
\begin{center}
\includegraphics[width=9cm]{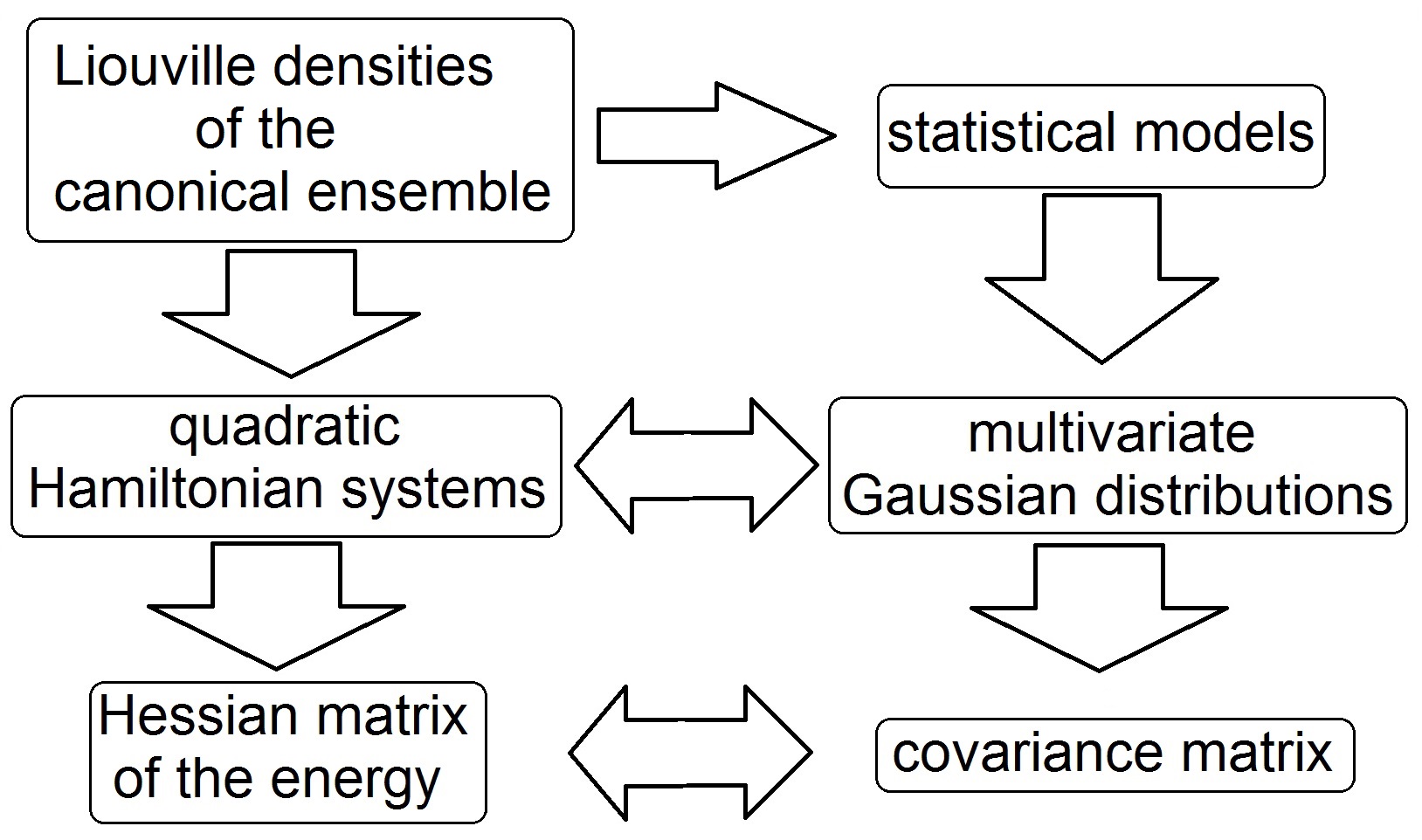}
\end{center}
  \caption{The Liouville density of the canonical ensemble defines a family of PDFs parameterized by macroscopic quantities which constitutes a statistical model. For the special case of the quadratic Hamiltonians, the correspondence between the Liouville densities and the multivariate Gaussian distributions is one--to--one and therefore also between the Hessian energy and the covariance matrices respectively.}\label{fig:quad}\end{figure}
\begin{figure}[!!ht]
\begin{center}
\includegraphics[width=11cm]{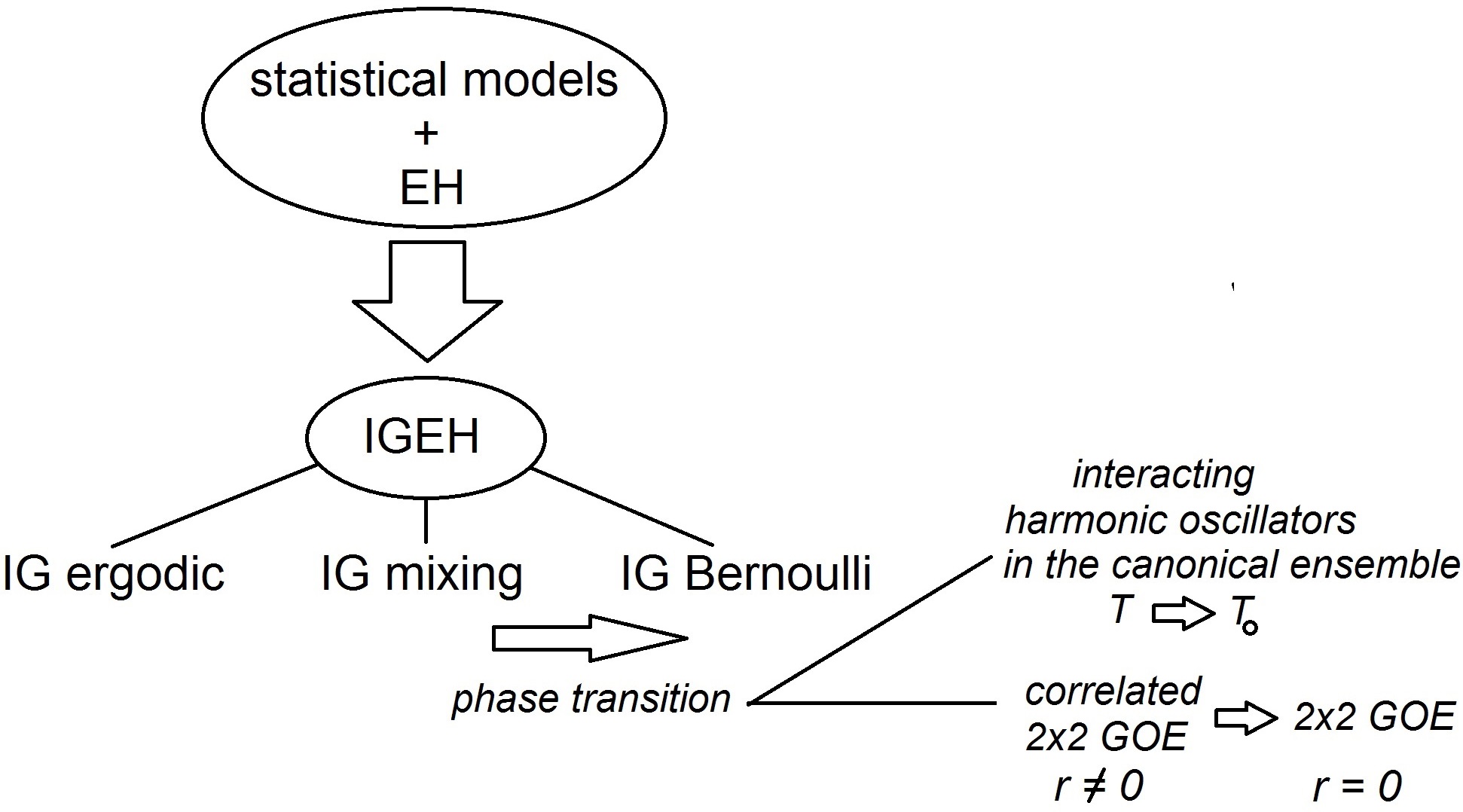}
\end{center}
  \caption{A diagram showing the way in which the IGEH can characterize phase transitions in terms of the passage from one of their levels to other, and in different applications: a pair of interacting harmonic oscillators in the canonical ensemble and the $2\times 2$ GOE.}\label{fig:IGEH}\end{figure}

Here we summarize the aspects of our proposal that can provide innovative tools within the context of the information geometry for characterizing the dynamics of a system in terms of the macrospace of the chosen statistical model. For this, below we provide two schematic diagrams, Figs. \ref{fig:quad} and \ref{fig:IGEH}, showing the content of the two main contributions of this work and its physical relevance from a panoramic outlook.

\section{Conclusions}\label{sec conclusions}
We have proposed an extension of the ergodic, mixing and Bernoulli levels in the context of information geometry, that we called \emph{information geometric ergodic hierarchy (IGEH)}, and we applied it to characterize: i) the phase transitions of a pair of interacting harmonic oscillators in the canonical ensemble and ii) the $2\times 2$ Gaussian Orthogonal Ensembles. The relevance and novelty of our main contributions, i.e. the IGEH and the information geometric characterization of the Lioville densities of the canonical ensemble expressed by the Eqs. \eqref{3-4}--\eqref{3-5Bernoulli} and \eqref{canonical2}--\eqref{canonical12} respectively, lie in the following remarks:
\begin{itemize}
  \item Statistical models provides a unified scenario for approaches involving correlations between microscopic variables. This was illustrated with a correlated $2\times 2$ GOE by adding correlations between two variables and showing that, this modification attenuates the chaotic dynamics on macrospace by increasing the scalar curvature, accordingly to the Cafaro's criterium of global chaos.

  \item 
      The IGEH generalizes the chaos characterization of the ergodic hierarchy by quantifying the statistical independence between the microvariables (instead of subsets of phase space) of the statistical model. This is performed in the asymptotic limit of large values of the time--like parameter which is expressed in terms of upper bounds of the IG correlation as the measure $F(p)$ for the case of the $2D$ correlated model.


  \item The association between multivariate Gaussian distributions and
quadratic Hamiltonians can be useful for studying the type of stability that present the dynamics in their equilibrium points in the context of the information geometry.

\item Geometrical notions and the Cafaro's criterium of global chaos can be related with the levels of the IGEH. The $2\times 2$ GOE case belonging to the most chaotic level, the IG Bernoulli, has an associated minimum negative value of the scalar curvature $R_{GOE}=-\frac{1}{2}$.
\item By obtaining upper bounds $F(p)$ on the IG correlation for a specific family of probability distributions, as exemplified by the curve of Fig. \ref{fig:distinguible}, one could study geometrical phase transitions moving along curves $F(p)$ as an external parameter $r$ is varied.
\end{itemize}



\section*{Acknowledgments}
This work was partially supported by CONICET and Universidad Nacional de La Plata, Argentina.

\appendix
\section{Proof of the distinguishability measure $F(p)$ (formula \eqref{4-3})}
\begin{proof}
Replacing \eqref{2Dmodel} in the definition of $F(p)$ one obtains
\begin{eqnarray}\label{8.3-4}
&F(p)=\nonumber\\
&\max_{\{(x,y)\in \mathbb{R}^2\}}\left|\frac{1}{2\pi\Sigma^2\sqrt{1-r^2}}\exp\left(\frac{-1}{2(1-r^2)}\left[\frac{(x-\mu_x)^2}{\sigma^2}+\frac{y^2\sigma^2}{\Sigma^4}-\frac{2r(x-\mu_x)y}{\Sigma^2}\right]\right)-
\frac{1}{\sqrt{2\pi\sigma^2}}\exp\left(-\frac{(x-\mu_x)^2}{2\sigma^2}\right)\frac{\sigma}{\sqrt{2\pi\Sigma^4}}\exp\left(-\frac{y^2\sigma^2}{2\Sigma^4}\right)\right| \nonumber
\end{eqnarray}
Now, by defining the following adimensional variables $\widetilde{x}=\frac{x-\mu_x}{\sigma}$ and $\widetilde{y}=y\frac{\sigma}{\Sigma^2}$ one can recast $F(p)$ as
\begin{eqnarray}\label{8.3-5}
F(p)=\frac{1}{2\pi\Sigma^2}m\acute{a}x_{\{(\widetilde{x},\widetilde{y})\in \mathbb{R}^2\}}\left|\frac{1}{\sqrt{1-r^2}}\exp\left(\frac{-1}{2(1-r^2)}\left(\widetilde{x}^2+\widetilde{y}^2-2r\widetilde{x}\widetilde{y}\right)\right)-
\exp\left(-\frac{\widetilde{x}^2+\widetilde{y}^2}{2}\right)\right|
\end{eqnarray}
Therefore, in order to calculate $F(p)$, it is enough to study the maximum and minimum on $\mathbb{R}^2$ of the function  $G_r(x,y)=\frac{1}{\sqrt{1-r^2}}\exp\left(\frac{-1}{2(1-r^2)}\left(x^2+y^2-2rxy\right)\right)-
\exp\left(-\frac{x^2+y^2}{2}\right)$ for each value of the parameter $r\in[-1,1]$. For this, it is convenient to write $G_r(x,y)=g_1(x,y)-g_2(x,y)$ with $g_1(x,y)=\frac{1}{\sqrt{1-r^2}}\exp\left(\frac{-1}{2(1-r^2)}\left(x^2+y^2-2rxy\right)\right)$ and
$g_2(x,y)=\exp\left(-\frac{x^2+y^2}{2}\right)$.
Then, one must to find the critical points of $G_r(x,y)$ from the equations
\begin{eqnarray}\label{8.3-6}
&\frac{\partial G_r(x,y)}{\partial x}=\frac{\partial g_1(x,y)}{\partial x}-\frac{\partial g_2(x,y)}{\partial x}=
g_1(x,y)\left(\frac{-1(x-ry)}{(1-r^2)}\right)-g_2(x,y)(-x)=0 \\
&\frac{\partial G_r(x,y)}{\partial y}=\frac{\partial g_1(x,y)}{\partial y}-\frac{\partial g_2(x,y)}{\partial y}=
g_1(x,y)\left(\frac{-1(y-rx)}{(1-r^2)}\right)-g_2(x,y)(-y)=0 \nonumber
\end{eqnarray}
From \eqref{8.3-6} it follows that $y\frac{\partial G_r(x,y)}{\partial x}-x\frac{\partial G_r(x,y)}{\partial y}=0$, thus
\begin{eqnarray}\label{8.3-7}
yg_1(x,y)\left(\frac{-1(x-ry)}{(1-r^2)}\right)+g_2(x,y)xy-
xg_1(x,y)\left(\frac{-1(y-rx)}{(1-r^2)}\right)-g_2(x,y)xy=0 \nonumber
\end{eqnarray}
Since $g_1(x,y)$ is an exponential then is always positive. Then one obtains
\begin{eqnarray}\label{8.3-8}
g_1(x,y)\left(\frac{-y(x-ry)+x(y-rx)}{(1-r^2)}\right)=0 \ \ \Longrightarrow \ \ g_1(x,y)\left(\frac{r(y^2-x^2)}{(1-r^2)}\right)=0
\ \ \Longrightarrow r=0 \ \ \textrm{or} \ \ y=\pm x \nonumber
\end{eqnarray}
If $r=0$ then it is clear that $p$ is equal to the product of their marginals and it follows that the infinite norm is zero. Assuming $r\neq0$
and replacing $y=\pm x$ in the formula of $G_r(x,y)$ then one obtains a function that only depends on $x$, that we can denote as $\varphi_r(x)$. Then one has that
\begin{eqnarray}\label{8.3-9}
\varphi_r(x)=\frac{1}{\sqrt{1-r^2}}\exp\left(\frac{-x^2}{(1-r^2)}\left(1\mp r\right)\right)-
\exp\left(-x^2\right)
\end{eqnarray}
Thus, the value of $F(p)$ is given by the maximum value of $|\varphi_r(x)|$ with $x\in\mathbb{R}$. For this, we calculate the critical points of
$\varphi_r(x)$ by making its derivative equal to zero, thus obtaining
\begin{eqnarray}\label{8.3-10}
&\frac{d\varphi_r(x)}{dx}=\frac{1}{\sqrt{1-r^2}}\exp\left(\frac{-x^2}{(1-r^2)}\left(1\mp r\right)\right)
\left(\frac{-2x}{(1-r^2)}\left(1\mp r\right)\right)+2x\exp\left(-x^2\right)=
2x\left[\exp\left(-x^2\right)-\frac{1}{\sqrt{1-r^2}}\exp\left(\frac{-x^2}{(1-r^2)}\left(1\mp r\right)\right)
\left(\frac{1}{(1-r^2)}\left(1\mp r\right)\right)\right]=0 \nonumber\\
&\Longrightarrow \ \ x=0 \ \ \ \textrm{or} \ \ \ \exp\left(-x^2\right)-\frac{1}{\sqrt{1-r^2}}\exp\left(\frac{-x^2}{(1-r^2)}\left(1\mp r\right)\right)
\left(\frac{1}{(1-r^2)}\left(1\mp r\right)\right)=0 \nonumber
\end{eqnarray}
So the critical points of $\varphi_r(x)$ satisfy
\begin{eqnarray}\label{8.3-11}
x=0 \ \ \textrm{or} \ \ \frac{1}{\sqrt{1-r^2}}\exp\left(\frac{-x^2}{(1-r^2)}\left(1\mp r\right)\right)
\left(\frac{1}{(1-r^2)}\left(1\mp r\right)\right)=\exp\left(-x^2\right)
\end{eqnarray}
Replacing \eqref{8.3-11} in \eqref{8.3-9} one has the value of $\varphi_r(x)$ in its critical points $x_c$
\begin{eqnarray}\label{8.3-12}
\varphi_r(x_c)=\left\{\varphi(0)=\frac{1}{\sqrt{1-r^2}}-1 \ \ ,  \ \pm r \exp\left(-x_{c}^2\right) \right\}
\end{eqnarray}
In order to find explicitly $\exp\left(-x_{c}^2\right)$ one must solve the second relation of the Eq. \eqref{8.3-11}, taking the natural logarithm in both sides of it then one has
\begin{eqnarray}\label{8.3-13}
-\frac{x_c^2(1\mp r)}{(1-r^2)}+\ln \left(\frac{1}{\sqrt{1-r^2}(1-r^2)}(1\mp r)\right)=-x_c^2 \nonumber
\end{eqnarray}
That is,
\begin{eqnarray}\label{8.3-14}
\exp\left(-x_c^2\right)=\left(\frac{1}{\sqrt{1-r^2}(1\pm r)}\right)^{\left(\frac{1\pm r}{\mp r}\right)}
\end{eqnarray}
Then, from \eqref{8.3-14} and \eqref{8.3-12} one obtains the following values of $\varphi_r(x_c)$ in its critical points
\begin{eqnarray}\label{8.3-15}
r\left(\sqrt{1-r^2}(1+r)\right)^{-1-\frac{1}{r}} \ \ , \ \ -r\left(\sqrt{1-r^2}(1-r)\right)^{-1+\frac{1}{r}}
\ \ , \ \ \frac{1}{\sqrt{1-r^2}}-1
\end{eqnarray}
Now, given that the maximum and minimum values of $G(x,y)$ are precisely the same of $\varphi_r(x)$ subject to the restriction $y=\pm x$ and since the maximum value of $|G(x,y)|$ is equal to $F(p)$, then from \eqref{8.3-15} one deduces that $F(p)$ is
\begin{eqnarray}\label{8.3-16}
F(p)=\max \left\{
|r|\left(\sqrt{1-r^2}(1+r)\right)^{-1-\frac{1}{r}},|r|\left(\sqrt{1-r^2}(1-r)\right)^{-1+\frac{1}{r}},\frac{1}{\sqrt{1-r^2}}-1 \right\}
\end{eqnarray}
from which follows the desired result.
\end{proof}

\end{document}